%% file: main_IJHCI.tex
\documentclass[]{interact}

\usepackage{epstopdf}
\usepackage[caption=false]{subfig}

\usepackage[natbibapa,nodoi]{apacite}
\renewcommand\bibliographytypesize{\fontsize{10}{12}\selectfont}

\usepackage{siunitx}
\usepackage{multirow}

\usepackage{graphicx}
\usepackage{hyperref}
\usepackage{stackengine}

\theoremstyle{plain}

\theoremstyle{definition}

\theoremstyle{remark}

\usepackage{cleveref}
\usepackage{xspace}
\usepackage{tabularx}
\usepackage{booktabs}

\newif\ifdraft
\draftfalse

\usepackage{xcolor}

\newcommand{\devicename}{TofuML\xspace}
\ifdraft
    \newcommand{\key}[1]{\textcolor{red}{{Key: #1}}}
    \newcommand{\new}[1]{\textcolor{red}{#1}}
    \newcommand{\todo}[1]{\textcolor{red}{\textbf{TODO: #1}}}
    \newcommand{\sugano}[1]{\textcolor{orange}{{[Sugano: #1]}}}
    \newcommand{\kawabe}[1]{\textcolor{magenta}{{[Kawabe: #1]}}}
    \newcommand{\nakao}[1]{\textcolor{blue}{{[Nakao: #1]}}}
    \newcommand{\shitara}[1]{\textcolor{green}{{[Shitara: #1]}}}
\else
    \newcommand{\key}[1]{}
    \newcommand{\new}[1]{#1}
    \newcommand{\todo}[1]{}
    \newcommand{\sugano}[1]{}
    \newcommand{\kawabe}[1]{}
    \newcommand{\nakao}[1]{}
    \newcommand{\shitara}[1]{}
\fi

\begin{document}


\title{\devicename: A Spatio-Physical Interactive Machine Learning Device for Interactive Exploration of Machine Learning for Novices}

\author{
\name{Wataru Kawabe\textsuperscript{a}\thanks{CONTACT Wataru Kawabe. Email: wkawabe@ut-vision.org}, Hiroto Fukuda\textsuperscript{a}, Akihisa Shitara\textsuperscript{b}, Yuri Nakao\textsuperscript{c}, and Yusuke Sugano\textsuperscript{a}}
\affil{\textsuperscript{a}The University of Tokyo, Tokyo, Japan; 
\textsuperscript{b}Tsukuba University, Tsukuba, Japan;
\textsuperscript{c}Fujitsu Limited, Kawasaki, Japan;
}
}

\maketitle

\begin{abstract}
We introduce TofuML, an interactive system designed to make machine learning (ML) concepts more accessible and engaging for non-expert users. 
Unlike conventional GUI-based systems, TofuML employs a physical and spatial interface consisting of a small device and a paper mat, allowing users to train and evaluate sound classification models through intuitive, toy-like interactions. 
Through two user studies---a comparative study against a GUI-based version and a public event deployment---we investigated how TofuML impacts users' engagement in the ML model creation process, their ability to provide appropriate training data, and their conception of potential applications.
Our results indicated that TofuML enhanced user engagement compared to a GUI while lowering barriers for non-experts to engage with ML.
Users demonstrated creativity in conceiving diverse ML applications, revealing opportunities to optimize between conceptual understanding and user engagement.
These findings contribute to developing interactive ML systems/frameworks designed for a wide range of users.
\end{abstract}

\begin{figure*}
  \centering
  \includegraphics[width=\linewidth]{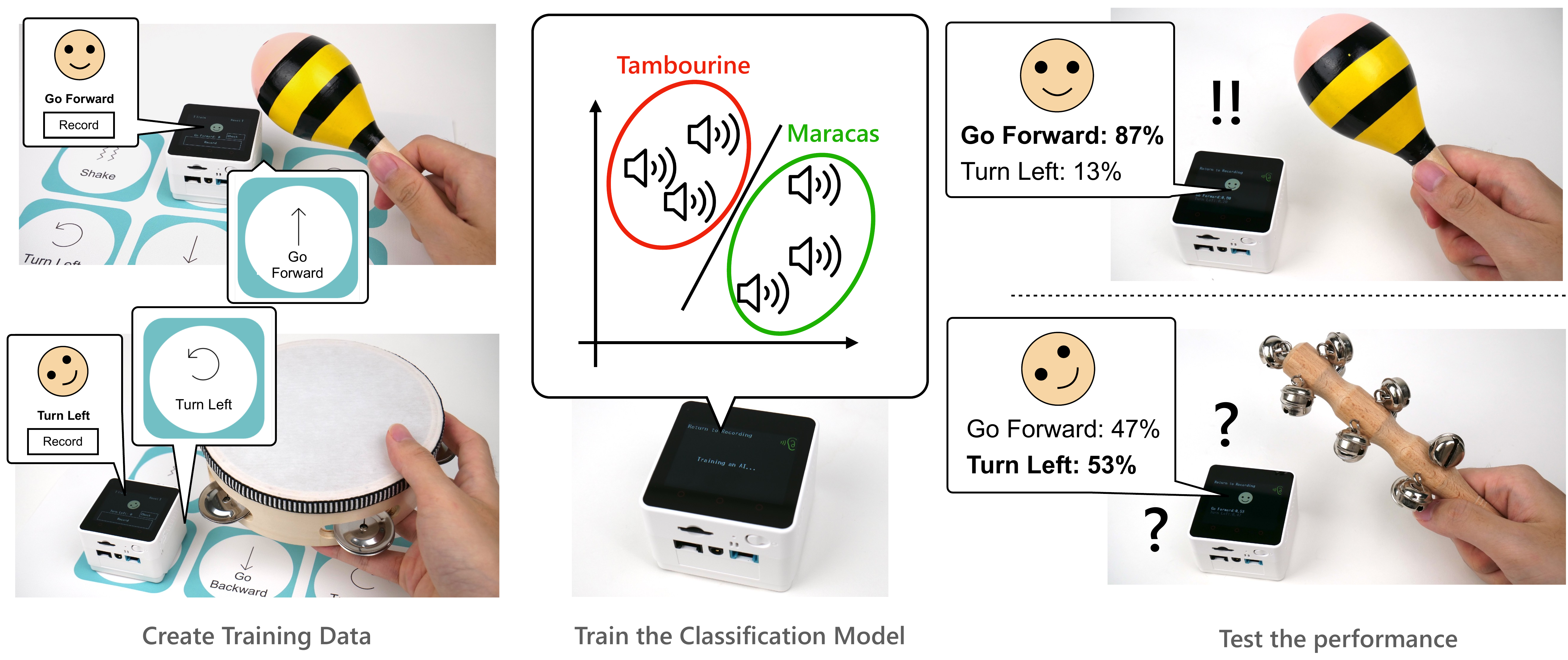}
  \caption{In this paper, we propose \devicename, an interactive machine learning (IML) system based on spatial and physical interaction to explore ways in which general users can engage more actively with the core concepts of machine learning (ML).}
  \label{fig:teaser}
\end{figure*}

\begin{keywords}
Interactive Machine Learning; Human-Centered Computing
\end{keywords}

\input{IJHCI/01_introduction.tex}

\input{IJHCI/02_relatedwork.tex}

\input{IJHCI/03_tangiblesystem.tex}
\input{IJHCI/04_guitui.tex}

\input{IJHCI/05_publicstudy.tex}
\input{IJHCI/06_results.tex}

\input{IJHCI/07_discussion.tex}
\input{IJHCI/08_conclusion.tex}

\section*{Acknowledgements}
This work was supported by JST SPRING Grant Number JPMJSP2108 and JST CREST Grant Number JPMJCR19F2. 
We would like to thank Miraikan, the National Museum of Emerging Science and Innovation (especially Hikaruko Aoki, Akira Ohkubo, Takahide Kato, Mayumi Sakuraba, and Chisa Mitsuhashi) for their generous support and cooperation. 
We also thank Akichika Tanaka, the designer of xMod and toio, for his warm support and guidance on hardware handling and software implementation.
We utilized OpenAI's ChatGPT for language editing and manuscript refinement.

\section*{Declaration of Interest Statement}
The authors have no competing interests to disclose.

\bibliographystyle{apacite}
\bibliography{main}


\end{document}

%% file: IJHCI/01_introduction.tex
\section{Introduction}

Machine learning (ML) has become an essential part of our daily lives, yet creating ML models remains a complex task that is accessible primarily to those with technical expertise. 
While large language models (LLMs)-based systems like ChatGPT have democratized certain aspects of AI, there remains a significant need for users to create customized, lightweight ML models for specific applications, particularly in resource-constrained or IoT environments. 
To make the process of creating ML models more approachable for users, physical interactions that resemble familiar objects or toys could provide a more intuitive entry point than traditional GUIs.
This paper introduces \devicename, a novel spatial-physical interface that transforms the core ML process (data collection, training, and inference) into an intuitive, tangible experience through a unique combination of autonomous mobility and position-based interaction.

Traditional Interactive Machine Learning (IML) systems have focused on making ML development more accessible through graphical user interfaces~\citep{dudley2018review,fails2003interactive,amershi2014power}. 
However, these GUI-based approaches often maintain a conceptual distance between users and the underlying ML mechanisms.
While some research has explored physical interfaces for ML~\citep{kaspersen2021machine,hitron2019can}, these systems have not fully investigated how spatial awareness and physical embodiment can create more engaging and intuitive ML experiences.

\devicename presents a new technical approach to IML that leverages spatial interaction and physical embodiment. 
The system consists of a small autonomous device that combines on-device ML capabilities with precise positioning, and a paper-based interaction space that maps ML functions to physical locations.
This design creates several technical challenges:
\begin{itemize}
\item Implementing efficient, real-time ML training and inference on a resource-constrained mobile platform.
\item Creating a robust spatial interaction system that maintains accuracy while allowing natural movement.
\item Developing an interaction model that balances technical sophistication with intuitive understanding.
\end{itemize}

Our system addresses these challenges through several key technical innovations. 
The autonomous device integrates sensors, processing capabilities, and output mechanisms in a compact form factor, enabling real-time sound processing and classification. 
The paper mat creates defined interaction zones that map to specific ML functions, allowing users to ``program'' the device's behavior through physical positioning.
This spatial-physical interaction creates a direct, tangible connection between user actions and ML concepts while maintaining the technical accuracy necessary for effective model creation.

To evaluate the effectiveness of our approach, we conducted two complementary studies.
First, we performed a controlled comparison between \devicename and a GUI-based version of the same functionality. 
This study revealed that while the GUI version offered certain usability advantages, \devicename significantly increased user engagement and led to more creative exploration of ML concepts. 
Second, we conducted a public study where general users interacted with the system in an unstructured environment. 
This study demonstrated how \devicename's physical interface naturally encouraged experimental use by non-experts and helped users develop an intuitive understanding of ML concepts.

The key technical contributions of this work include:
\begin{itemize}
\item A novel spatial-physical interaction method for hands-on ML exploration that combines autonomous mobility with position-based function mapping.
\item A complete hardware/software architecture that enables mobile, embodied ML interaction while maintaining real-time responsiveness.
\item Implementation techniques for creating interactive ML experiences on resource-constrained mobile platforms.
\item Empirical insights into the effectiveness of physical embodiment in ML interfaces, including comparative analysis with GUI-based approaches.
\end{itemize}
Through our analysis of user interactions and feedback, we demonstrate how \devicename's technical innovations create new opportunities for non-expert users to engage with the ML concept.
The system's ability to combine technical sophistication with intuitive physical interaction suggests promising directions for future research in embodied ML interfaces.

%% file: IJHCI/02_relatedwork.tex
\section{Related Work}

\subsection{IML and User Analysis}

Interactive machine learning (IML) is a framework that enables users to efficiently and effectively interact with ML systems to develop ML models through appropriate UI design~\citep{dudley2018review,fails2003interactive,amershi2014power}. 
IML typically assumes that users train ML models using locally stored data or data collected on the spot and aims to improve the ease and explainability of this process.
\new{The simplest examples are systems and games where users can train ML models using images and audio that they capture on the spot~\citep{carney2020teachable,webcampacman,fiebrink2010wekinator,mellis2017machine}.}
Various approaches have been proposed to support users in the IML process.
For example, these include visualizations to understand the model's state~\citep{talbot2009ensemblematrix,zhou2016making,zhang2018manifold}, functionalities that allow users to construct the ML development process itself~\citep{franccoise2021marcelle,mishra2021designing}, and frameworks that use user-provided feedback as training data to improve the model~\citep{xin2018accelerating,arendt2019towards,ramos2020interactive,shin2025looping}.
These approaches can be considered enhancements that allow users with strong motivation for solving tasks and a clear understanding of what kind of data and models are appropriate for effectively achieving their goals.

Research focuses on analyzing users' behaviors and thoughts during IML experiences to gain insights for better system design rather than directly proposing improved IML systems. 
Some examples have organized events to explore challenges in designing robust ML systems and investigate user reactions to develop new methodologies~\citep{weitz2024explaining,nakao2020use,kawabe2024technical,hitron2019can,yang2018grounding,goodman2021toward}.
For example, Kawabe et al.~\citep{kawabe2024technical} examined whether users could acquire the necessary technical understanding for ML concepts through IML experiences.
Nakao and Sugano~\citep{nakao2020use} studied how non-expert deaf and hard-of-hearing individuals could deepen their understanding of ML-based sound recognition systems.
Hitron et al.~\citep{hitron2019can} used a pre/post experimental design to investigate whether children could construct essential mental models related to ML. 
These studies, which explore users' understanding and perceptions after IML experiences, aim to realize more user-friendly and effective IML systems by gaining deeper insights into user interactions and cognition.

IML research has advanced along two main paths: system proposals and user analysis. 
These studies typically focus on enabling users to create more accurate models efficiently or promoting technical understanding among those already interested in ML. 
However, how to increase engagement among diverse users who may initially lack interest in ML has not been sufficiently explored. 
We are particularly interested in IML systems that can spark users' spontaneous interest and foster the generation of diverse ideas. 
We propose an IML system based on physical and spatial interactions as one possibility to achieve this. 
We verify its effectiveness by comparing it with conventional IML systems and observing users' spontaneous interactions.

\subsection{Physical and Spatial ML Experiences}

Expanding from ML model design, real-world ML applications have been studied in diverse fields, with examples using physical and tangible~\citep{ishii1997tangible} devices.
Enabling ML experiences through devices can provide users with more immersive experiences or offer ML interactions that involve real-world environments. 
For instance, by incorporating ML models into robots, users can experience ML inference while feeling as if they are interacting with humans~\citep{semeraro2023human,sheridan2016human}. 
Additionally, embedding ML models in wearable accessories such as smartwatches allows for more seamless integration of ML into everyday scenarios~\citep{jain2022soundwatch,mujawar2023smart,nahavandi2022application}.
Moreover, there are cases where everyday objects, such as a pen or glasses, are transformed into ML-enabled devices, aiming to make our daily activities more efficient~\citep{matulic2020pensight,lin2020empathics,murase2012gesture}.
While these applications highlight the transformative impact of ML across various device types, they essentially remain within the realm of pre-trained model applications.
Our research focuses on the training process itself, enabling users to actively participate in creating and refining ML models rather than merely utilizing pre-trained ones.

In ML model prototyping, leveraging real-world devices as interactive triggers has proven effective in overcoming technical barriers. 
Kaspersen et al. explored this concept by introducing a box-shaped device that accepts handwritten inputs, making the ML process more tangible and accessible~\citep{kaspersen2021machine}. 
Similarly, Hitron et al. integrated a tangible input device with a graphical user interface to facilitate the interactive design of a gesture recognition model, enhancing user engagement and feedback~\citep{hitron2019can}. 
In an educational context, Williams et al. introduced PopBots, a social robot that allows children to use a tablet to train an ML model, customizing the robot's responses in real-time~\citep{williams2019artificial}. 
Additionally, Kim and Kwon examined the effectiveness of an AI curriculum for children incorporating tangible computing tools, demonstrating its potential to enrich learning experiences~\citep{kim2024tangible}.
These systems are implemented with the expectation that introducing devices would have a positive impact on users' ML development experience.
Our system's novelty lies in expressing output through the device's own actions and in investigating the impact of such physical presentations on users.

To provide users with a more immersive ML experience, besides embedding ML models into devices or robots, dynamically utilizing space can also be considered.
Bakogeorge et al., for example, proposed a tabletop IML system designed to foster trust and collaboration between multiple users through synchronous fine-tuning~\citep{bakogeorge2024embodied}.
Even without using physical space, virtual reality can be utilized to provide ML experiences such as data collection for hand gesture classification~\citep{bahceci2022supervised}, medical diagnosis through ML-based analysis of user task performance~\citep{belger2023application}, and immersive exhibition viewing leveraging ML~\citep{chang2021digital}.
Furthermore, even on a GUI, it is possible to create training data and train models using two-dimensional space~\citep{chang2021spatial,tatsuya2020investigating}.
While previous research primarily uses spatial interactions for data exploration, our approach with \devicename uniquely allows users to define the device's output actions through spatial positioning. 
This enables a more direct and intuitive engagement with ML concepts by physically mapping input sounds to output actions in space.

Previous studies have investigated users' ability to develop appropriate ML models and the new perspectives acquired through these experiences, showing similarities to our research. 
However, our study's uniqueness lies in its focus on a diverse group of general users who did not primarily participate for experimental purposes. 
When recruiting participants in advance, there's a potential bias toward users with shared technical backgrounds. 
Our study minimizes this bias by hosting events in public spaces and recruiting participants on-site, thereby capturing a more diverse and representative sample of users.

\subsection{Interactive Sound Recognition}

In this paper, we propose \devicename, a system that allows users to interactively train and evaluate sound recognition models. 
This system offers an experience where users can personalize a sound recognizer according to their own objectives. 
Various use cases for personalized sound recognizers have been investigated and proposed.
Notable examples include studies on the usage of commercial voice interaction-based speech recognizers~\citep{lopatovska2019talk,berdasco2019user} like Amazon Alexa~\citep{amazonAlexa} and Apple Siri~\citep{siri}. 
Additionally, there are cases where deaf or hard-of-hearing users create their own recognizers by registering sound sources relevant to their daily lives~\citep{goodman2021toward,jain2022protosound,jain2022soundwatch}.
Other scenarios include AI agents that provide feedback on pronunciation improvements in language learning~\citep{liu2022ai} and personalized music emotion recognizers~\citep{wang2012personalized}. 
These examples demonstrate various approaches to adapting models to individual data for solving user-specific tasks.

Our study also investigates the process of users creating personalized sound recognizers. 
While previous research has focused on developing models that directly address everyday challenges, our research examines the significance of experiencing \devicename as a first step in ML prototyping for users.
\devicename itself is not designed to solve any specific task. 
This flexibility may enable a wide range of potential use cases as users envision. 
We hypothesized that this open-ended approach would allow users to connect ML model solutions to problems they conceive.
By providing an unrestricted platform, we aim to observe how users naturally associate ML capabilities with their own ideas and challenges.

%% file: IJHCI/03_tangiblesystem.tex
\section{\devicename: A Spatio-Physical IML Device}

The primary focus of this study is to consider and develop an IML system that various non-expert users can use, introduce the created prototype to ML experience events, and evaluate its effectiveness there.

\subsection{System Overview}

We designed an interactive IML system with the concept that a diverse range of users, including those without experience with electronic devices, could easily try the core ML process as if they were playing with a toy. 
The task we focused on was sound classification. 
We chose classification as one of the most fundamental tasks in ML, making it suitable for introducing users to the model creation experience.
We focus on sound compared to other media, considering the advantage that producing various sounds for training can be done more intuitively and immersively.

While the IML development flow includes various detailed steps such as model selection and quality assessment~\citep{dudley2018review}, we restricted the functionalities to the minimum necessary for creating an ML model.
The steps for the users are limited to three essential steps: providing training data, training the model with the provided data, and evaluating the performance of the trained model.
Our system skips advanced steps essential to ML development, such as hyperparameter tuning, model architecture design, and data augmentation.
This streamlined approach focuses on the ML model creation's core elements while simplifying the user process.


\begin{figure}[t]
  \centering
  \includegraphics[width=0.8\linewidth]{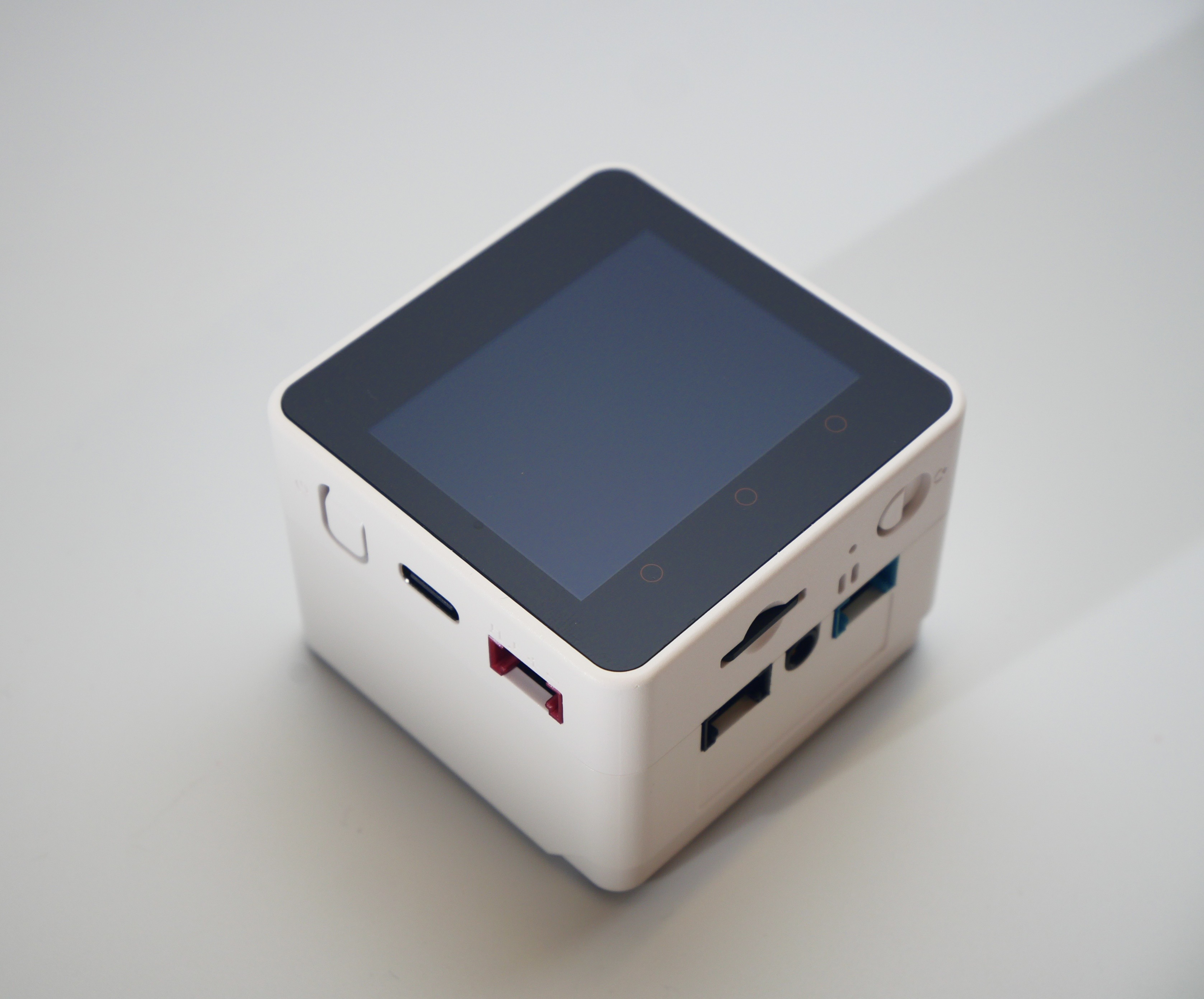}
  \caption{An overview of \devicename. Through the manipulation of this cube-shaped device, users can experience training and evaluating a sound classification model.}
  \label{fig:device_overview}
\end{figure}

We created a more user-friendly device than PCs, tablets, or smartphones to develop an IML system that reduces reliance on traditional GUI interactions while improving user engagement.
Figure~\ref{fig:device_overview} shows the appearance of the device. 
To achieve familiarity, we adopted a simple, hand-held cuboid shape, giving the device a toy-like appearance that users can easily manipulate. 
In addition to displaying numerical probabilities, the device expresses the model's output results through various physical actions, such as wheel movement, body vibration, and LED lights.
This approach provides a more intuitive and tangible representation of the model's decisions than, e.g., showing category probabilities on displays.

\begin{figure}[t]
  \centering
  \includegraphics[width=0.6\linewidth]{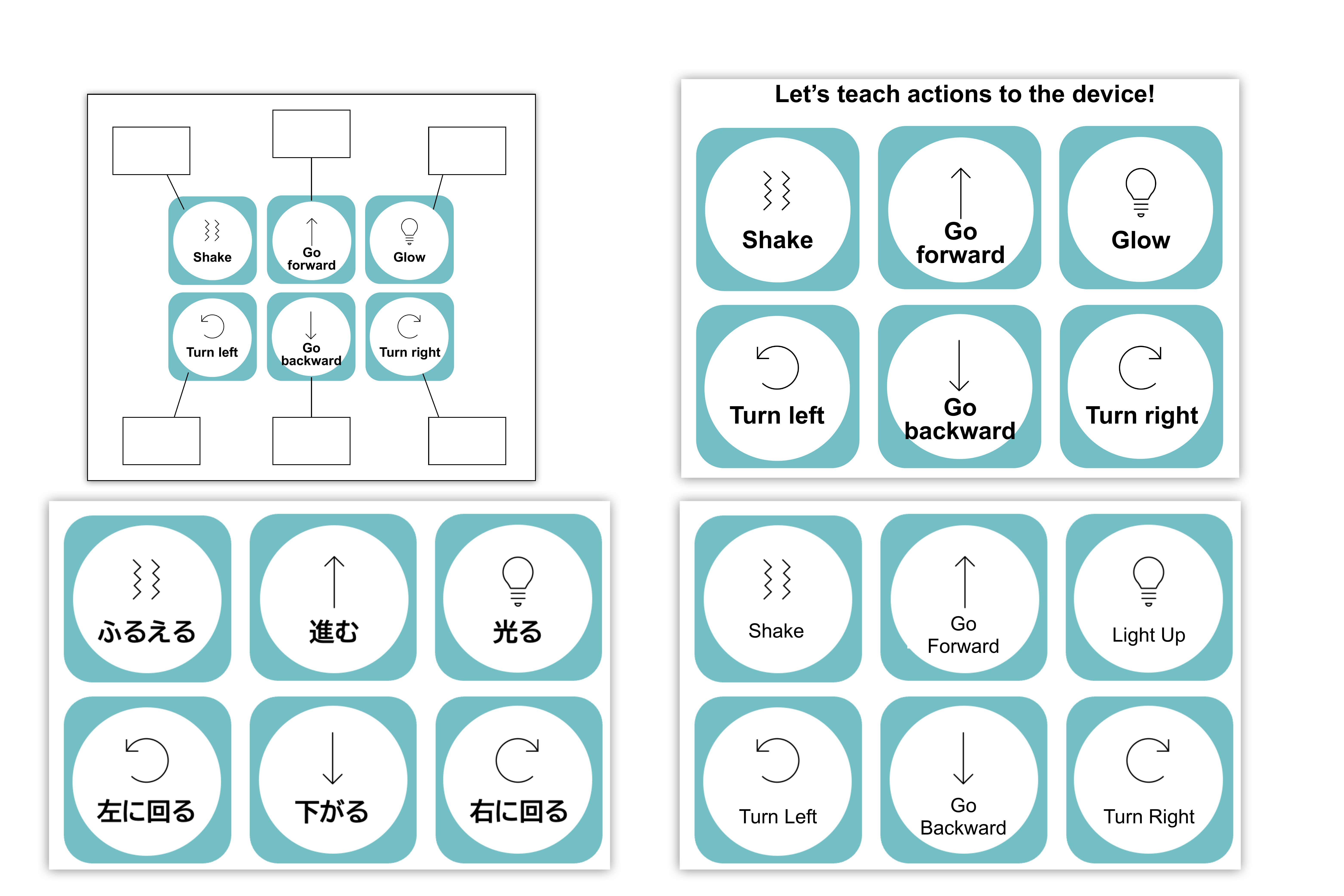}
  \caption{A paper mat for associating recorded sounds with specific actions in the training data. When users place the device in each area, they can create training data with the output corresponding to the action associated with that area.}
  \label{fig:mat}
\end{figure}

As described earlier, our device's primary purpose is to create a classification model that maps a specific sound to a physical output action.
The device captures sound through its microphone as input for the sound classification model and outputs the actions.
We designed a paper mat, as shown in Fig.~\ref{fig:mat}, in addition to the device. 
To enhance usability while reducing cognitive load, we adopted an approach that allows the user to specify the device's output by changing its position on the mat.
This mat features basic actions (\textit{shake}, \textit{go forward}, \textit{light up}, \textit{turn left}, \textit{go backward}, and \textit{turn right}), each corresponding area to an action the device can perform.
Users can record training data corresponding to the device's output action by placing it in a specific position on the mat.
This setup allows users to create training data and interact with the model in a tangible, intuitive manner.

\subsection{Interaction Flow}



\begin{figure*}[t]
  \centering
  \includegraphics[width=.98\linewidth]{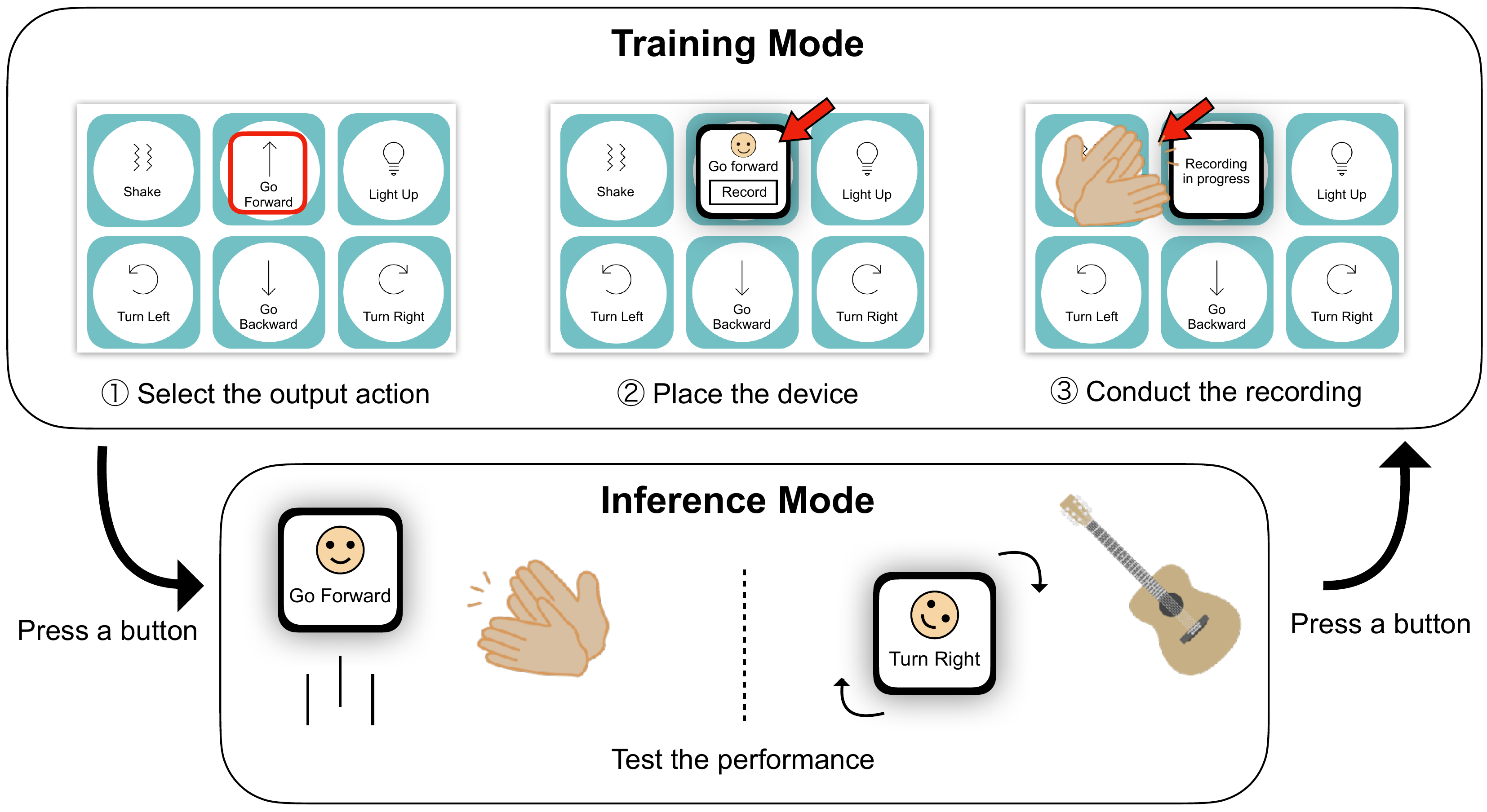}
  \caption{The interaction flow of \devicename. First, in the training mode, users decide on the action they want the device to perform and place it in the corresponding area. The recorded sound is linked to the output action and added to the training data by recording in this state. After training, the device moves on to the inference mode by doing this for multiple actions. In the inference mode, it differentiates its actions based on the sounds it hears.}
  \label{fig:flow}
\end{figure*}

Figure~\ref{fig:flow} illustrates the basic interaction flow for creating and evaluating a sound classification model using \devicename.
The workflow begins with a training mode, where the user decides on sound-action pairs they want the device to learn. 
For example, they might want to train the device to move forward when they clap their hands and turn right when they speak.
To create training data, the user places the device on a specific area on the mat, which corresponds to a certain action.
The device's screen updates to display the selected action name and an emoticon. 
The user then presses the record button on the screen and produces the corresponding sound. 
This process is repeated to create multiple samples for each action, with the sample count displayed on the screen.

Once sufficient data is collected for all desired actions, the user initiates training by pressing a physical button on the device. 
The device automatically transitions into inference mode once the model's training is complete.
In this mode, the device records sounds from the microphone at regular intervals, classifies them, and performs the action with the highest probability.
During inference, the user can test the model by producing various sounds. 

If the model's performance is unsatisfactory, they can return to the training mode by pressing the physical button. 
This allows for iterative improvement by adding more samples or retraining the model as needed.
The screen displays detailed probability distributions for each action during inference for debugging purposes. 
The system also includes features for managing training data, such as deleting the most recent sample or all samples if a reset is necessary.

\subsection{Implementation Details}

\begin{figure}[t]
  \centering
  \includegraphics[width=0.7\linewidth]{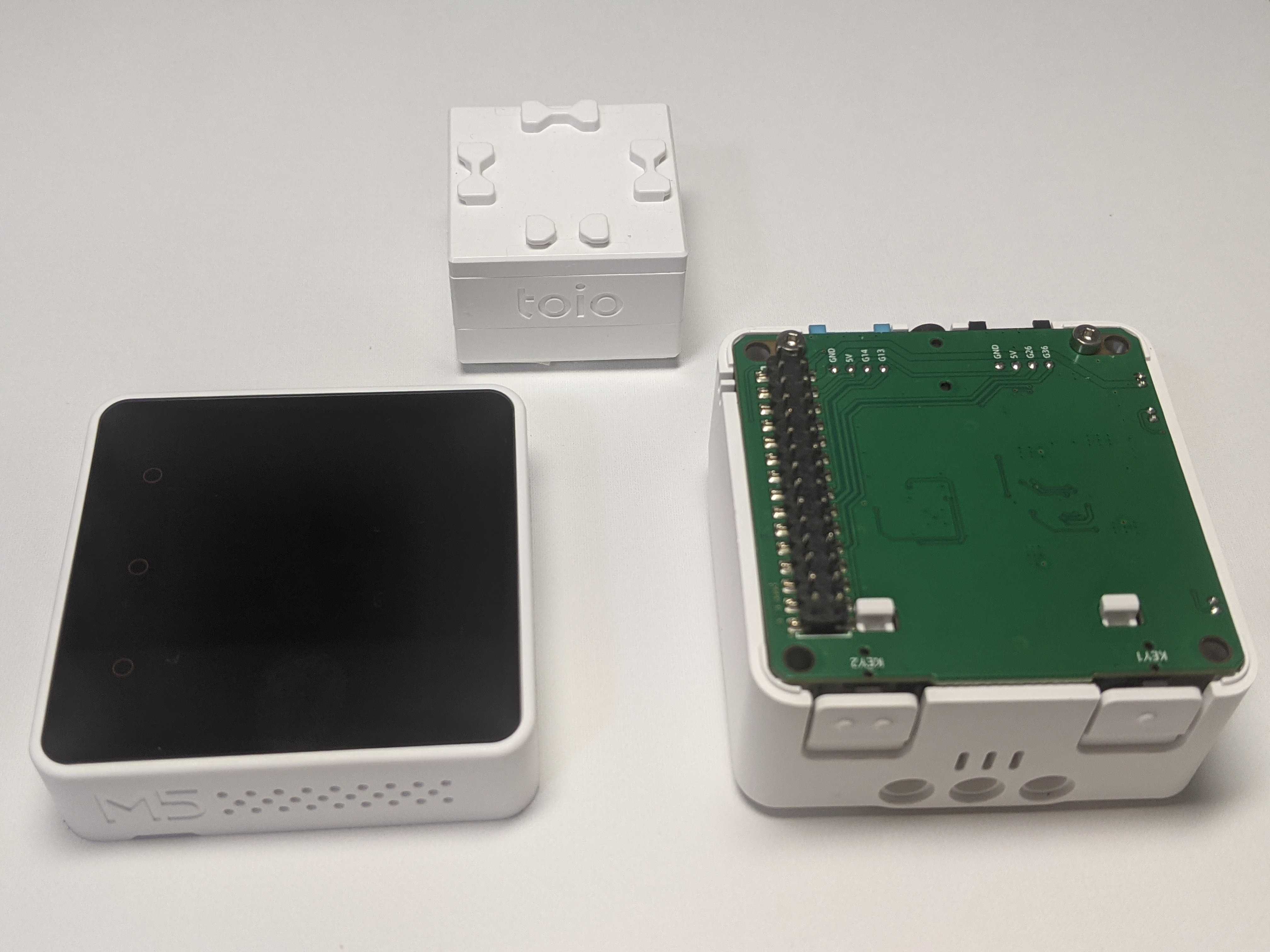}
  \caption{M5Stack Core2 (right), toio (middle), and the attached base part (left). The M5Stack Core2 and the base part are pre-assembled, and users utilize the device by embedding the toio into the base and integrating the two components.}
  \label{fig:device_separate}
\end{figure}

For the main hardware, we primarily used M5Stack Core2\footnote{\url{https://docs.m5stack.com/ja/core/core2}}, an Arduino-compatible device equipped with a small touch screen and network I/O capabilities.
We embedded a toio\footnote{\url{https://www.sony.com/en/SonyInfo/design/stories/toio/}}, which has wheels, into a base part attached to M5Stack Core2 to enable physical movement (Fig.~\ref{fig:device_separate}).
This base part, which we developed, features two physical buttons used to operate the device.
The M5Stack Core2's bus is connected to various components in the base part: speakers (including the LRA) through an amplifier, and the microphone, expansion terminal, and MIDI through a GPIO expansion IC. 
The amplifier is specifically connected to the speakers and LRA, while the other components are linked via the GPIO expansion IC.
M5Stack Core2 and toio are connected via Bluetooth; any command to move the wheels is sent to the toio. 
The underside of the toio is equipped with an optical sensor that recognizes an invisible special pattern printed on the mat.
We used a mat compatible with toio and designed the layout ourselves.
This allows the toio to determine its absolute position on the mat and identify which area it is currently located in.
Among these hardware components, M5Stack Core2 and toio are off-the-shelf products, while we developed the base part ourselves.

The ML algorithm operates separately on a local server, and the device communicates with this server through a Wi-Fi network.
During training, the sounds recorded by users and the associated action names are sent to the server.
Each sound recorded by users is one second long, and these individual samples are sent to the server along with the ID of the associated output action.
During inference, the periodically recorded sounds are transmitted.
The inference results are then returned to the device and reproduced as a physical action.
During inference, the device records for one second at 2.5-second intervals. 
The delay between the completion of recording and the execution of the action output by inference is approximately 3 seconds.

We followed the implementation of the sound classification model in a previous study~\citep{kawabe2024technical}.
The model extracts sound features from a convolutional neural network pre-trained with AudioSet~\citep{hershey2017cnn,gemmeke2017audio}.
The classification algorithm is the Random Forest~\citep{breiman2001random} which is trained using the extracted features.
The feature extraction process follows the official implementation of the TensorFlow sample code.
It is lightweight enough to perform training on a CPU.
We chose this model architecture for its ease of use, which ensures that the training finishes without making users wait for a minute.

%% file: IJHCI/04_guitui.tex
\section{A Comparative Study with GUI}

We developed \devicename due to the observation that existing GUI-based IML systems do not necessarily increase engagement for general users.
We conducted an in-laboratory evaluation to compare the usability and user engagement of \devicename with GUI-based approaches. 
This study aimed to determine which system is more effective in stimulating users' interest in ML model creation and potentially increasing engagement.

\subsection{GUI Implementation}

\begin{figure*}[t]
    \centering
    \subfloat[Training mode. Users click each action button to start recording. The digit in each action button shows the number of samples users have recorded so far. \label{fig:baseline_1}]{%
        \includegraphics[width=0.48\textwidth]{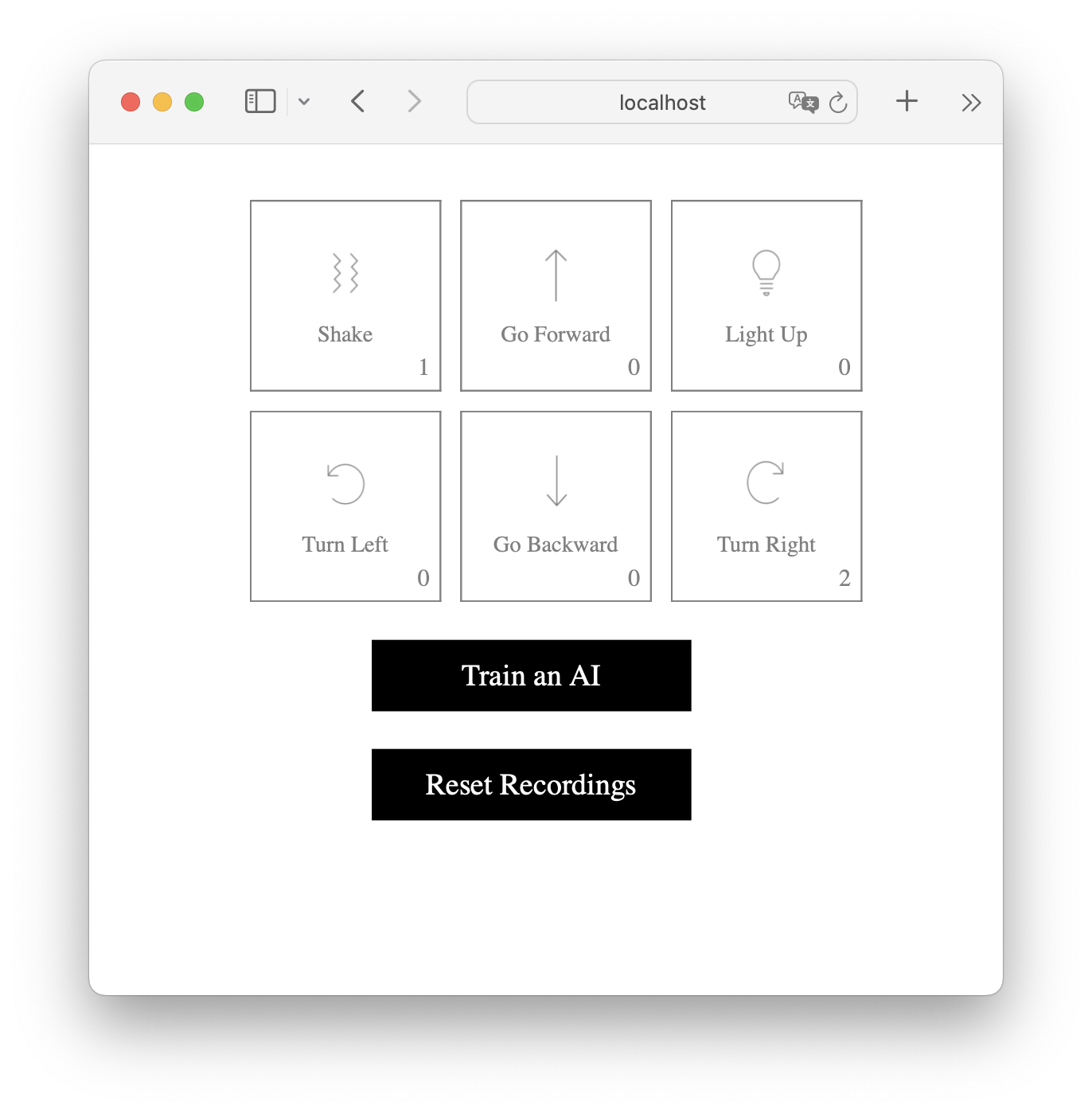}%
    }
    \hfill
    \subfloat[Inference mode. The microphone captures sound at regular intervals, and the model performs inference. The probability for each action is displayed and the action with the highest probability is highlighted with a darker color and an icon at the bottom of the screen. \label{fig:baseline_2}]{%
        \includegraphics[width=0.48\textwidth]{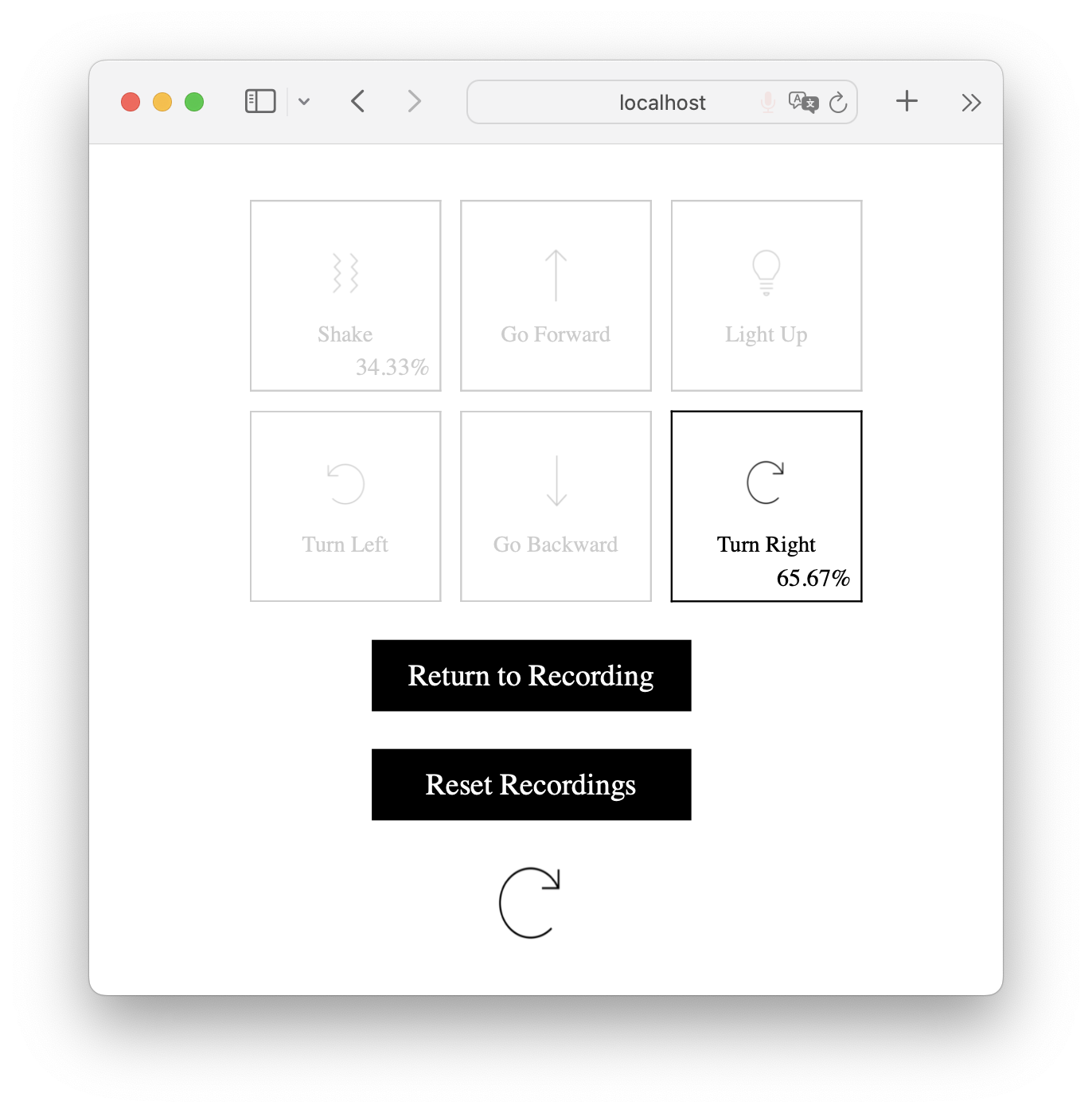}%
    }
    \caption{The overview of the GUI version of the IML system.}
    \label{fig:baseline}
\end{figure*}

To compare the effects of \devicename and its GUI version, we re-implemented the basic functionalities of the \devicename using basic GUI components.
We created a web application that aimed to be as similar as possible to the physical version regarding the UI overview, user capabilities, and information presentation methods.
For example, we adopted the same sound classification model and training method as \devicename and standardized the recording duration timing.
We believe this approach would allow participants to focus on comparing the two interfaces, enabling us to gather more detailed opinions on the benefits and drawbacks of the GUI and \devicename.

We show the overview of the GUI version in Fig.~\ref{fig:baseline}.
Like the paper mat of \devicename, six different actions are arranged on the screen (Fig.~\ref{fig:baseline_1}). 
We implemented each area as a button corresponding to placing the device and recording in \devicename. 
First, in the training mode, users can start recording by clicking these buttons and reset all recordings by clicking the ``Reset recordings'' button if necessary.
Just as the number of recordings is displayed on the device's screen, the recording count is shown in the bottom right corner of each button.
When the user clicks the ``Train an AI'' button, the model is trained on the user-created training data and automatically transitions to inference mode after training. 
In the inference mode, the model captures sound through the PC's microphone at regular intervals and performs inference.
The action with the highest probability in the inference is highlighted with a darker color, and its icon is displayed at the bottom of the screen, as shown in Fig.~\ref{fig:baseline_2}. 
Additionally, the probability values are shown within each action button.
During inference, users can return to the training mode by clicking the ``Return to Recording'' button, allowing them to create new recordings.


\subsection{Procedure}

Before commencing the study, we informed our university's ethics review committee about the study content and obtained permission to conduct it.
We recruited participants through university mailing lists and online bulletin boards.
Each participant visited our laboratory space individually and signed a consent form after explaining the study.
They were instructed to compare and evaluate two IML systems objectively, assuming we conduct an AI/ML-related event for end users.
Given the limitations of human sound production, we provided various sound-making goods, such as maracas, tambourines, and glockenspiels, for participants to use freely.
Participants developed sound classification models through the systems using a set of goods. 
We counterbalanced the order of the systems the participants experienced.
Participants experienced each system for a minimum of 10 minutes and a maximum of 15 minutes. 
After experiencing both systems, they completed an online questionnaire and participated in an interview to conclude the session.
\new{
To replicate the use cases envisioned by the system, we allowed participants to define classification tasks. 
Consequently, it is practically impossible to establish test data for quantitative performance evaluation for each participant. 
Therefore, the performance assessment of the models created by participants was excluded from the scope of this experiment.
}

Our evaluation focuses on whether users can increase their engagement and show proactivity when introduced to \devicename or its GUI version as an opportunity to experience AI/ML model creation. 
To analyze which aspects of TofuML contributed to user engagement, we developed custom questionnaire items that addressed both specific system features and general usability aspects rather than using existing standard usability scoring methods.
\new{Following Lalmas et al.'s definition of user engagement as a multi-dimensional construct incorporating emotional aspects~\citep{lalmas2022measuring}, we developed the following eight items:}
\begin{description}
    \item[Q1] Overall enjoyment of the experience
    \item[Q2] Perception of the system's design as attractive
    \item[Q3] Enjoyment of the recording process
    \item[Q4] Pleasure in observing the output results (i.e., actions)
    \item[Q5] Enjoyment of the trial-and-error process
    \item[Q6] Sense of achievement when the trained model works
    \item[Q7] Contribution to learning about AI and ML
    \item[Q8] Ease of use of the system
\end{description}

We employed a pairwise comparison method using a 5-point scale, as adopted in the previous study~\citep{law2019comparing}. 
\new{We adopted pairwise comparison rather than Likert scale ratings because participants might hesitate to give low scores and rate both systems favorably. 
Pairwise comparison encourages participants to directly evaluate the relative strengths and weaknesses of the two systems.}
Participants were presented with a linear scale for each question, with five equidistant circles representing the response options. 
The leftmost end was labeled ``PC version'' (i.e., the GUI version), while the rightmost end was labeled ``Toy version'' (i.e., \devicename). 
For analysis purposes, we assigned numerical values to these responses, ranging from -2 (preferring GUI) to +2 (preferring \devicename), with 0 representing a neutral stance. 

We also included open-ended questions asking participants to describe both systems' positive and negative aspects and their thoughts when comparing the two. 
Considering that some participants might find it challenging to express their opinions fully in writing, we addressed the same points through interviews to ensure we captured a comprehensive range of responses.
\new{Two of the authors transcribed the interview recordings and conducted the coding analysis of the transcripts.}


\subsection{Results}

\begin{table}[t]
    \caption{Demographics of participants of the laboratory study.}
    \label{tab:demographics_guitui}
    \begin{tabular}{lccccc}
        \toprule
        AI/ML Experience & Male & Female & Other & N/A & Total \\
        \midrule
        None & 4 & 7 & 0 & 1 & 12 \\ 
        Studying AI/ML & 2 & 0 & 0 & 0 & 2 \\
        Developing AI/ML & 0 & 1 & 0 & 0 & 1 \\
        Both & 3 & 1 & 0 & 0 & 4  \\
        \midrule
        Total & 9 & 9 & 0 & 1 & 19 \\
        \bottomrule
    \end{tabular}

\end{table}

We collected data from a total of 19 participants \new{with ages} ranging from 19 to 53 (M $=32.63$, SD $=10.84$), and the breakdown is shown in Tab.~\ref{tab:demographics_guitui}. 
12 participants had no previous experience learning or developing AI or ML. 
Five participants reported having development experience.

\begin{figure*}[t]
  \centering
  \includegraphics[width=\linewidth]{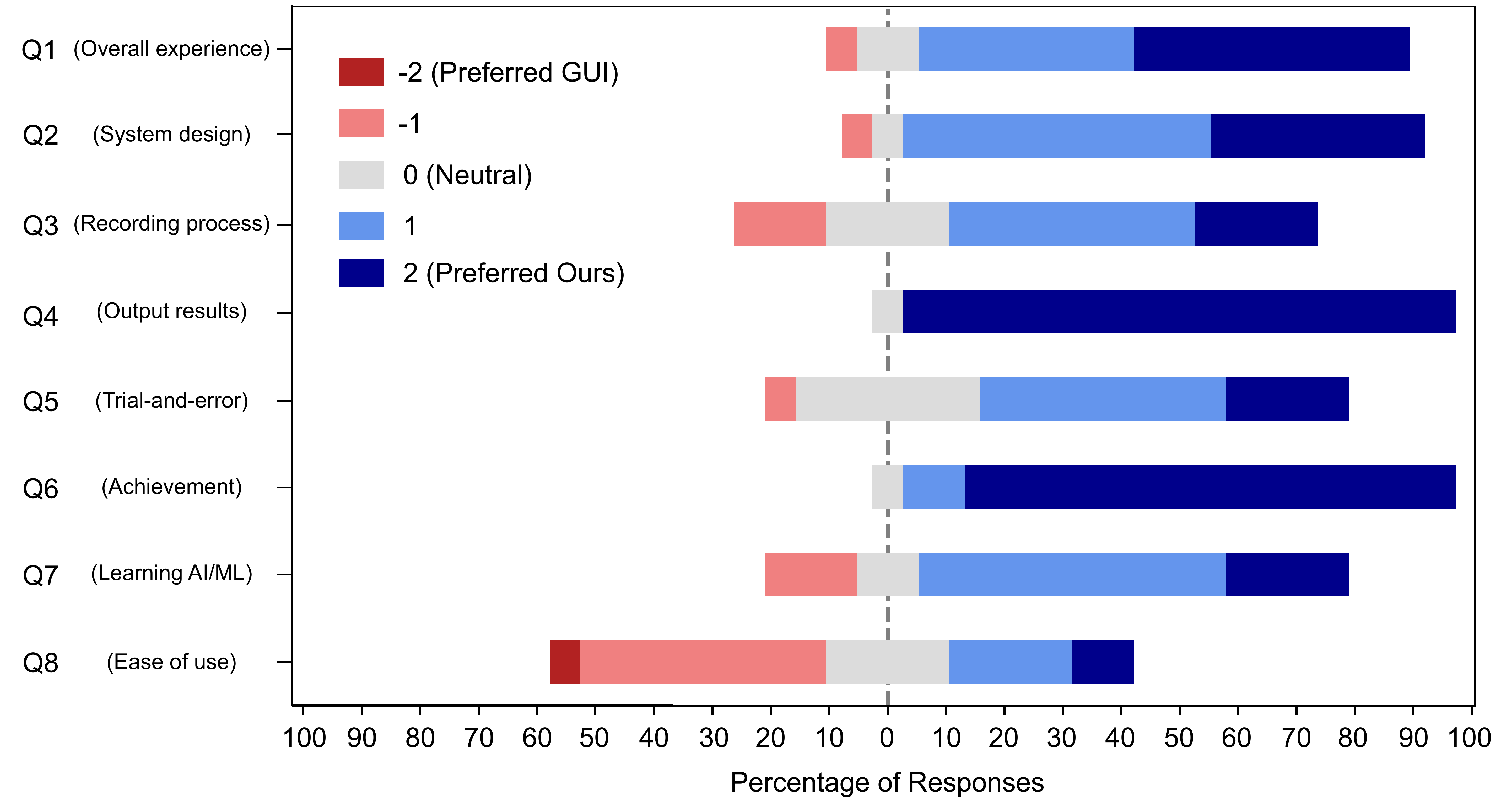}
  \caption{Responses of pairwise comparison between \devicename and the GUI version.}
  \label{fig:comparative_guitui}
\end{figure*}

\subsubsection{Quantitative Results}
\label{result1}
Our pairwise comparison results, presented in Fig.~\ref{fig:comparative_guitui}, reveal a clear preference for \devicename over the GUI version across multiple dimensions. 
The average scores for each item were as follows: Q1 scored $1.26$, Q2 scored $1.21$, Q3 scored $0.68$, Q4 scored $1.89$, Q5 scored $0.79$, Q6 scored $1.79$, Q7 scored $0.79$, and Q8 scored $-0.11$, with $0.00$ representing a neutral stance (equal preference for both systems). 
Except for Q8, which represents ``Ease of use of the system,'' all items preferred the \devicename. 
Notably, Q4 and Q6, which correspond to the pleasure of observing output and sense of achievement, respectively, stood out with particularly high scores.
These results support the conclusion that participants found \devicename more appealing and effective in various aspects of the ML experience than its GUI counterpart.

\subsubsection{Qualitative Results}
\label{result2}
When considering opinions on the benefits and drawbacks of the systems, we can observe many positive comments regarding \devicename, particularly concerning user engagement aspects such as approachability, cuteness, and enjoyment.
For example, P10 emphasized the strengths of \devicename as follows: ``\textit{Compared to the PC version, I felt a stronger sense of nurturing AI. 
The illustrations (facial expressions) and the device's size made it feel cuter and more approachable. 
I think it would be especially good for young children.}''
P1 found value in the device's ability to express inference results through movement: ``\textit{Since it actually moves, it's satisfying to see it respond correctly to sounds with the right movements. Also, when recording sounds, I feel more inclined to ensure the machine hears properly compared to when using a PC.}''
Furthermore, P8 preferred moving the device on the mat when creating training data: ``\textit{I thought it was good and easy to understand how to place the toy on the mat for recording. Also, since you can experience the toy moving, I felt that children, in particular, would find it enjoyable.}

Regarding the drawbacks of \devicename, several opinions were observed concerning the inconvenience caused by having to move the device and the device moving on its own.
P14 revealed their feelings when the model's performance was poor: ``\textit{With the PC version, even if it fails, I don't feel much because nothing moves. However, with the toy version, since it actually moves, I felt irritation and frustration towards the machine when it failed. It felt like common computer phenomena such as screen freezing or not working properly.}''
P9 pointed out, ``\textit{With the toy version, I worry it might fall or bump into things while I'm thinking 'oh no, oh no.' It might be good to place things around it that it can safely bump into.}'' 
The same participant, P9, also viewed the need to move the device itself as a disadvantage for creating training data: ``\textit{With the toy version, you have to move it to record and then move it again. But with the PC version, you can keep making sounds and clicking efficiently, so I think you can input (record) more efficiently.}''

Some participants found the GUI version more suitable for a comfortable experience. 
For instance, P5 appreciated the amount of information presented on the PC screen: ``\textit{It was good that I could see an overview of the recording counts for each movement. The output was also easy to understand as I could easily view the estimated probabilities at a glance.}''
There were also opinions that the GUI version's benefits were more apparent to those familiar with PC operations. 
P13 stated, ``\textit{The PC version was easy to use because I'm proficient with computers. The operation screen was simple and easy to understand.}'' 
P9 added, ``\textit{I think it's easy to operate as long as you can use a mouse.}''
Conversely, when compared to \devicename, many participants pointed out the lack of physicality and the impersonal nature of the GUI version as drawbacks. 
P1 stated, ``\textit{When simulating, for example, even if the 'Shake' option is triggered, it's disappointing that nothing actually shakes. I felt there was no sense of 'I made something move!'}''
P8 commented, ``\textit{While the recording and learning operations are straightforward, just clicking on the relevant parts of the same screen, it feels impersonal and somewhat lacks excitement.}''
These comments highlight that while the GUI version may be efficient and easy to understand, it lacks the tangible and engaging qualities that the physical \devicename provides.


%% file: IJHCI/05_publicstudy.tex
\section{A Public User Study}

\begin{figure*}[t]
  \centering
  \includegraphics[width=0.95\linewidth]{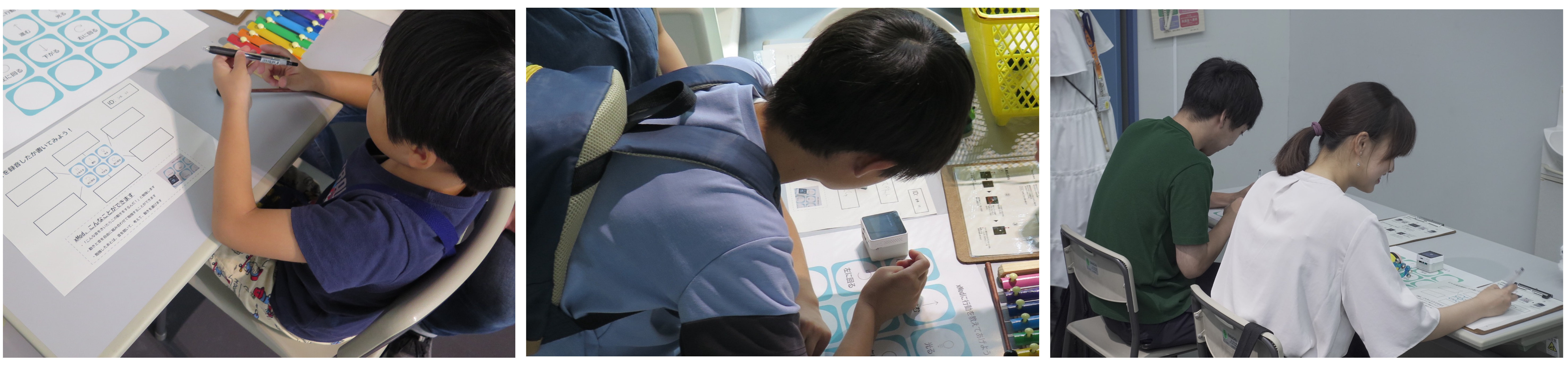}
  \caption{By analyzing users' experiences with \devicename, we comprehensively evaluated the system's impact on users. They autonomously created classification models using \devicename in the hands-on event, following their own goals. Through this flexible process, we investigated how they developed their ideas and what impressions they formed about ML creation process.}
  \label{fig:event_photo}
\end{figure*}

We additionally conducted a public study to more deeply investigate the reactions of non-expert users when \devicename, which emphasizes spatial and physical interaction, is introduced into their IML experience.
We aimed to investigate users' impressions, level of engagement, and generation of ideas about the ML implementation as they naturally interacted with our IML system.

\subsection{Study Overview}


Following an approach in a prior work~\citep{kawabe2024technical}, we organized an event and recruited attendees as study participants, conducting the study as a hands-on event.
Recruiting participants specifically for a study might result in a biased user group with technical backgrounds or strong preconceptions about the study's purpose. 
In contrast, visitors to public spaces are not prepared to participate in the study. 
We believed this would provide a sample of general users as close as possible to being unbiased.
We considered a science museum an appropriate venue for sampling the general public, as it is where families and friends visit to enjoy the exhibitions.
We did not make any prior announcements and set up the event so visitors could find the study on the spot and participate freely.
We designed our study to simulate a scenario where users autonomously operate the IML system to address their own challenges with an ML model. 
Rather than imposing quantitative goals, we encouraged participants to freely explore the system based on their self-defined objectives. 
To analyze their reactions through free-form experiences, we adopted a hackathon-style event~\citep{taylor2018everybody} format where participants created their own preferred ML models as part of the ML prototyping process (Fig.~\ref{fig:event_photo}).
We provided minimal instructions and interventions, allowing participants to engage with the device at their own pace without specific achievement targets. 
We structured the study as an open-ended experience, rather than a fixed-time event, where participants could come and go as they pleased. 
This approach allowed participants to expand their ideas and gradually enhance their technical understanding of the device, fostering a more personal and reflective exploration of how ML could be integrated into their daily lives and environments.


\subsection{Procedure}

\begin{figure}[t]
  \centering
  \includegraphics[width=0.8\linewidth]{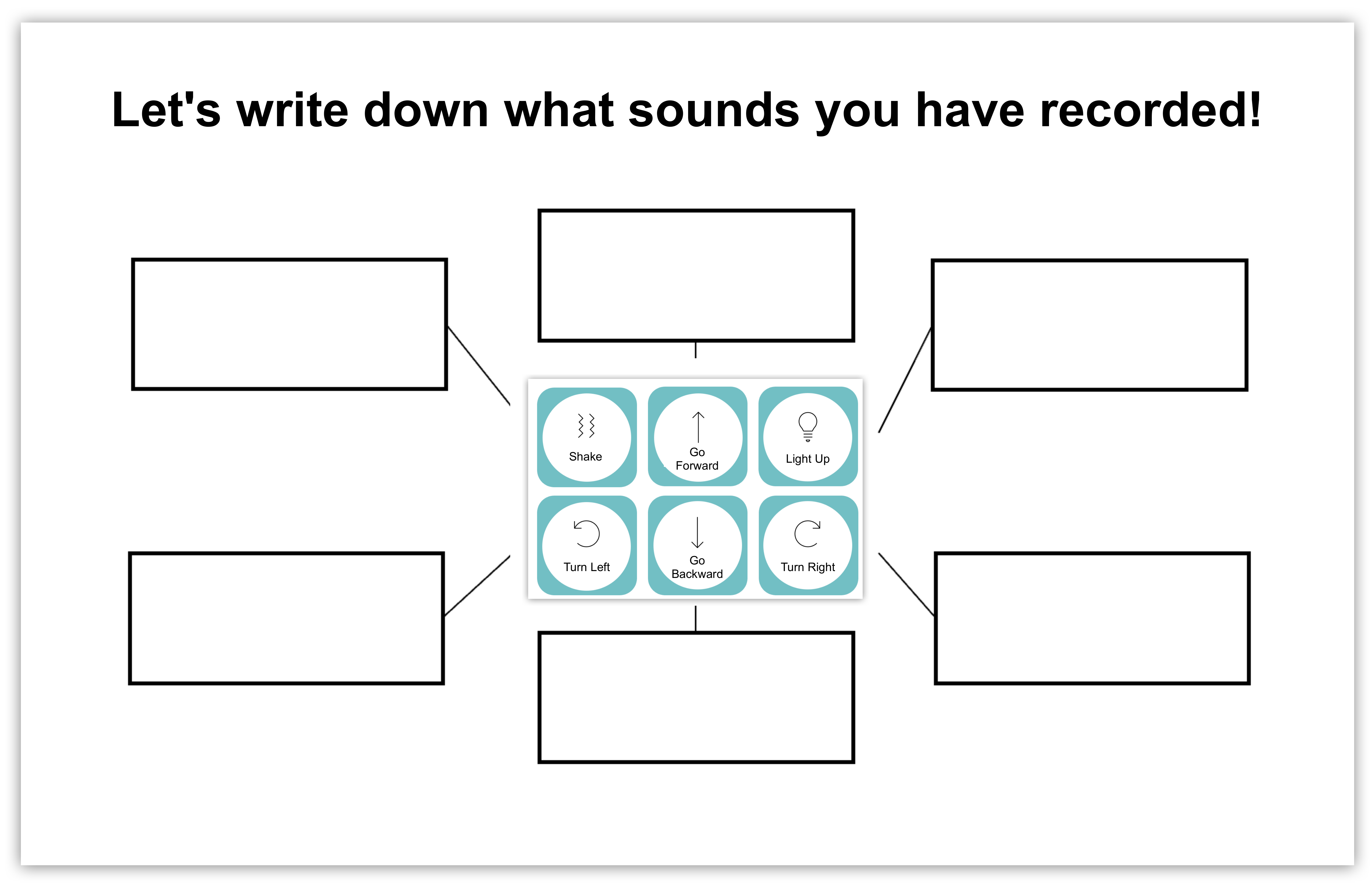}
  \caption{An overview of the worksheet for participants to note which sounds they associated with each action. We also utilized this as an analytical resource.}
  \label{fig:worksheet}
\end{figure}

Prior to commencing the study, we informed our university's ethics review committee about the study content and obtained permission.
The event took place in a glass-walled room on the exhibition floor of a science museum.
At the room's entrance, a device demonstration was shown, and visitors could freely observe and participate if they were interested.
In the demonstration, to ensure consistent information was conveyed to participants, we organized the items to be communicated beforehand and adhered to this prepared explanation.
After agreeing to participate, they received a quick explanation about the event and signed a consent form.
\new{For participants aged 12 or younger, parents reviewed the consent form and discussed its contents with their children before signing on their behalf. 
We also allowed parental support or intervention for these young participants.}
Participants were given a device, a paper mat, and a worksheet to note which sounds they associated with each action (see Fig.~\ref{fig:worksheet}).
Participants were then asked to select up to three items from various sound-producing goods similar to those used in the GUI comparison experiment and bring them to their tables. 
Once seated at their table, participants filled out the pre-experience questionnaire.
If participating as a group, one representative provided the responses.

Once the experience began, participants moved to tables outside the room if necessary to minimize the mixing of sounds from other participants. 
All the actions performed by the participants on the device were recorded for our analysis.
During the experience, staff members accompanied the participants to provide technical support and answer questions.
The staff's role was strictly technical support; participants had complete freedom to decide which sounds to associate with each action and record based on their preferences.
During the experience, participants took notes on the worksheet and continued the activity without any time restrictions until they were satisfied with their model.
After completing the experience, the participants filled out a post-experience questionnaire. 
If they consented, one of the authors conducted an in-depth interview with them to conclude the experience.

To comprehensively assess the impact of the \devicename experience, we collected data on participants' envisioned AI/ML use cases, their specific interests, and motivations during interaction, alongside their overall engagement.
We employed multiple response formats, including written and spoken, with short and long response options. 
This diverse approach ensured participants could express their thoughts fully and comfortably, accommodating individual preferences in communication styles.

\subsubsection{Questionnaire}

We investigated user engagement and ideas as AI/ML users and developers through a post-experience questionnaire. 
Items related to engagement and mental burden were measured using a 5-point Likert scale ranging from strongly disagree to strongly agree. 
\new{Rather than using generalized usability questions, our intention was to directly assess \devicename's effects on AI model creation. 
We designed custom question items specifically aligned with the context of model training and evaluation.}
The question items are shown below.
\begin{description}
    \item[Q1] You gained knowledge about AI.
    \item[Q2] You developed an interest in AI.
    \item[Q3] You felt the difficulty of creating AI.
    \item[Q4] You felt capable of creating AI.
    \item[Q5] You realized the importance of data for AI.
\end{description}
As open-ended questions, we asked the following items:
\begin{itemize}
    \item Please freely write about what you learned or understood through the experience.
    \item Please freely write about the good and bad aspects of the AI toy (i.e., the device) you used.
\end{itemize}
For the second item, we provided separate areas for writing the good and bad aspects.

We also examined participants' perspectives as users and developers both before and after the experience to assess the ML application ideas that emerged from it. 
All items were open-ended questions.
In the pre-experience questionnaire, we asked users, \textit{``How have you used AI in your daily life so far?''} and \textit{``Before this experience, have you ever learned about or created AI? If so, please describe the characteristics of the AI you created.''}
In the post-experience questionnaire, we asked users, \textit{``How do you think you can use AI in your daily life from now on?''} and \textit{``What kind of AI do you think you can create from now on?''}

\subsubsection{Interview}
In the interview, we used a semi-structured format to deeply understand the participants' thoughts and ideas. 
The questions focused on three main topics: 
\begin{itemize}
    \item What did you think about AI and the device through and after the experience?
    \item As an AI user, what use cases or application scenarios would you like or want in the future?
    \item As an AI designer, what use cases or application scenarios do you think you could design, or what would enhance your life if you created them yourself?
\end{itemize}
These questions aimed to elicit in-depth information during the conversation.

%% file: IJHCI/06_results.tex
\subsection{Results}  

\begin{table}[t]
    \caption{Demographics of participants who attended the public event and contributed to data collection.}
    \label{tab:demographics}
    \begin{tabular}{lccccc}
        \toprule
        AI/ML Experience & Male & Female & Other & N/A & Total \\
        \midrule
        None & 14 & 10 & 0 & 0 & 24 \\ 
        Studying AI/ML & 11 & 3 & 0 & 0 & 14 \\
        Developing AI/ML & 1 & 0 & 0 & 0 & 1 \\
        Both & 1 & 1 & 0 & 0 & 2 \\
        \midrule
        Total & 27 & 14 & 0 & 0 & 41 \\
        \bottomrule
    \end{tabular}

\end{table}

We collected data from 41 participants \new{with ages} ranging from 2 to 53 (M $=22.56$, SD $=15.28$), and the breakdown is shown in Tab.~\ref{tab:demographics}. 
24 participants had no previous experience learning or developing AI or ML. 
Three participants reported having development experience.
The average experience time was 1096 seconds, with a standard deviation of 581 seconds. 

\subsubsection{User Engagement}
\label{result3}

\begin{figure*}[t]
  \centering
  \includegraphics[width=\linewidth]{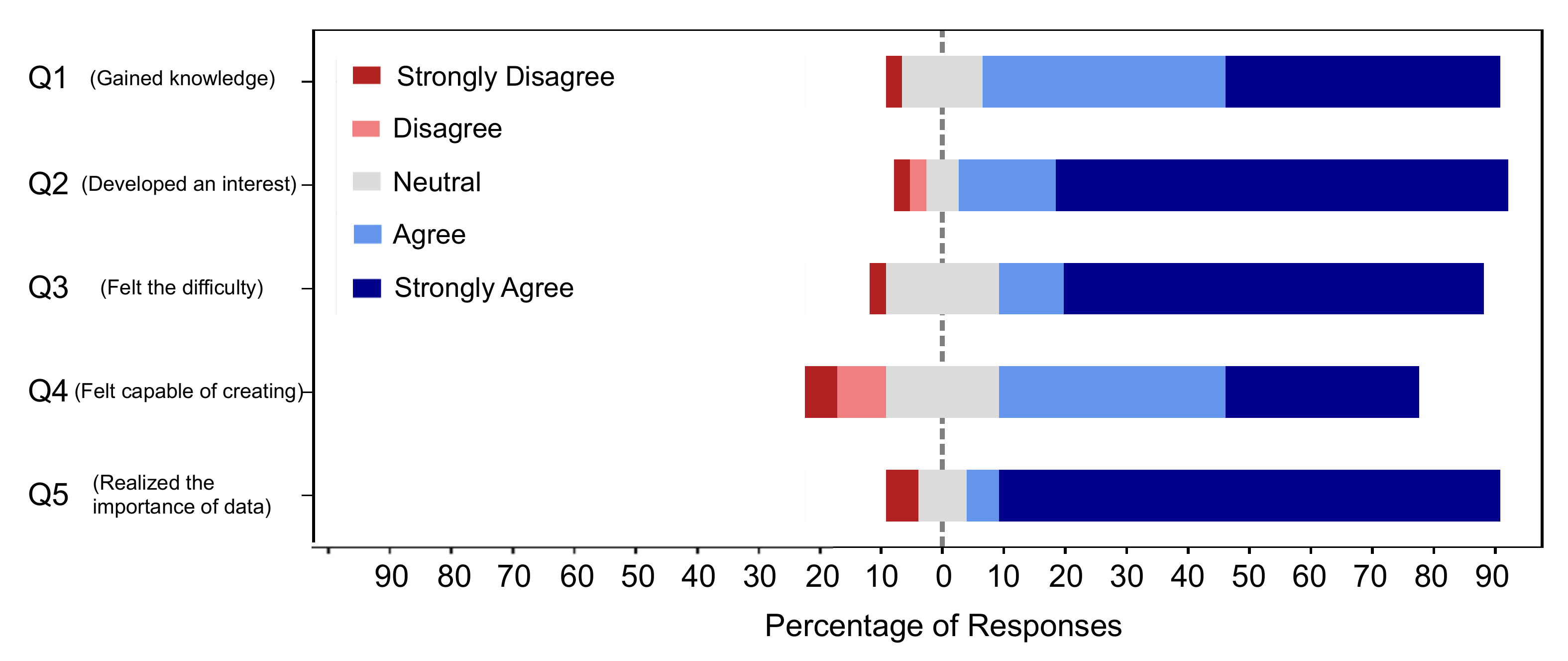}
  \caption{Responses to Likert scale questionnaire items related to user engagement and self-efficacy.}
  \label{fig:likert}
\end{figure*}

We show the details of the Likert scale questions in the questionnaire in Fig.~\ref{fig:likert}.
The participants' responses generally indicated high ratings across all items. 
This suggests that the participants enjoyed the overall experience, interpreted the model creation process in their own way, and became interested in ML model creation. 
Among all the items, Q4 received relatively lower scores. 
This can be attributed to the difficulty in improving the model's performance and the participants' awareness that they only engaged in a simplified development process.

We can also observe the positive feedback regarding the device design of \devicename. 
P6 stated, ``\textit{It was great that even children could easily experience AI. The operation was simple.}'' 
Similarly, P40 mentioned, ``\textit{It was easy to experience AI}'', and P10 remarked, ``\textit{It was easy to operate without special programming knowledge}''.
Regarding opinions about familiarity, P5 said, ``\textit{The expression on the screen was cute, and the accuracy improved with the number of recordings, which made me feel attached to it}''. P8 noted, ``\textit{I used to think AI was scary, but knowing it could be turned into a toy made it more approachable.}''
There were also comments expressing an analogy to living creatures or a sense of familiarity with the device, indicating increased user engagement brought about by physical UI. 
P13 stated, ``\textit{It is nice and convenient because it is a small machine.}'' 
P4 mentioned, ``\textit{It was fun to teach the AI (sounds),}'' and P34 said, ``\textit{It felt similar to a human},'' showing that some participants perceived the device as a small intelligent being.

On the other hand, negative feedback primarily stemmed from hardware-specific issues and the device's low explainability. 
Participants expressed various concerns about hardware problems. 
P6 noted that ``\textit{The vibration was difficult to perceive}'', while P1 commented, ``\textit{The position of the buttons was a bit unclear.}'' 
P19 pointed out a durability issue, stating, ``\textit{The parts came off easily.}''
Concerning the device's explainability, P36 said, ``\textit{I could not tell if it acted as I taught it}'', suggesting that displaying accuracy as a percentage alone is insufficient to understand the details of the performance. 
Additionally, P41's comment, ``\textit{I did not really understand how the AI works}'', indicated the challenge of evaluating whether the AI's behavior was correct.

\begin{table*}[t]
  \caption{The participants' responses to the pre-experience question: ``\textit{How have you used AI in your daily life so far?}''}
  \label{tab:idea_user_pre}
  \begin{tabularx}{\linewidth}{l X c}
    \toprule
    Category & Description & Responses \\
    \midrule
    Unclear or Non-usage & Responses indicating no or unclear usage & 12 \\
    Home \& Personal Assistance & Use of AI for music, alarms, and weather updates & 9 \\
    Education \& Learning & AI used in educational contexts and learning & 6 \\
    Lifestyle Assistance & AI used for daily assistance and interactions & 5 \\
    Creative Work & AI in creative tasks like image and text generation & 4 \\
    General AI Technology Mention & General mentions of using AI technologies & 4 \\
    Entertainment & AI used for explorative and fun activities & 1 \\
    \bottomrule
  \end{tabularx}
\end{table*}

\begin{table*}[t]
  \caption{The participants' responses to the post-experience question: ``\textit{How do you think you can use AI in your daily life from now on?}''}
  \label{tab:idea_user_post}
  \begin{tabularx}{\linewidth}{l X c}
    \toprule
    Category & Description & Responses \\
    \midrule
    Home \& Personal Assistance & AI to assist in daily chores and personal tasks & 10 \\
    General AI Utilization & Broad use of AI in everyday technology and automation & 6\\
    Healthcare \& Caregiving & AI for health diagnostics and caregiving & 5\\
    Education \& Learning  & AI used for educational support and children's learning & 3 \\
    Entertainment & AI for personal entertainment and leisure activities & 3\\
    Unclear or Non-specific & Ambiguous responses about AI capabilities & 3\\
    \bottomrule
  \end{tabularx}
\end{table*}

\begin{table*}[t]
  \caption{The participants' responses to the post-experience question: ``\textit{What kind of AI do you think you can create from now on?}''}
  \label{tab:idea_designer_post}
  \begin{tabularx}{\linewidth}{l X c}
    \toprule
    Category & Description & Responses \\
    \midrule
    Specific Needs-based AI & AI designed for specific tasks or needs & 11\\
    AI with Accessible Development & AI that can be developed with simple operations & 10\\
    Unclear or Negative Views & Unclear responses or skepticism about AI design& 4\\
    Conversational AI & AI that can engage in conversation or act as a companion & 3\\
    \bottomrule
  \end{tabularx}
\end{table*}

\subsubsection{Attitude towards AI}
\label{result4}

Based on pre and post-experience questionnaires, we investigated what ideas participants generated as AI/ML users and as developers/designers through their interactions with \devicename.
We summarized users' perspective before and after the experience in Tab.~\ref{tab:idea_user_pre} and Tab.~\ref{tab:idea_user_post}.
We also summarized the changes in participants' perspectives as developers/designers after the experience in Tab.~\ref{tab:idea_designer_post}.
Note that before the experience, only three participants had previous experience of developing AI/ML.

Regarding the use of AI, many participants already had experience using AI applications in their daily lives, with notable mentions of specific products such as smart speakers and ChatGPT. 
After the device experience, numerous application scenarios for lifestyle assistance were proposed, which seem irrelevant to the event content.
One notable point is that after the experience, participants' ideas shifted towards more concrete, data-driven applications of AI rather than vague notions of AI as a generically intelligent entity.
In the ``Home \& Lifestyle Assistance'' category, P28 mentioned, ``\textit{An AI that scans the refrigerator's contents to suggest possible recipes and identify items nearing their expiration dates}.''
P7 suggested, ``\textit{An AI that assesses the freshness of vegetables at the market based on camera image data. It would be useful if image data could be linked with freshness data.}'' 
These ideas reflect an awareness of acquiring performance through training using data, emerging from their experience of easily providing data for the device.

From the perspective of AI developers, only three participants had development experience before the experience. 
However, after the experience, various ideas were proposed, as shown in the Tab.~\ref{tab:idea_designer_post}.
One observed tendency is that the AI participants proposed to create from now on were not necessarily limited to the extensions of the experienced device, such as sound recognition or controlling the device. 
Specifically, some of the ideas included:
\begin{itemize}
    \item AI that alerts you to allergenic foods
    \item AI that allows non-verbal individuals to control household appliances through sounds
    \item AI that measures a person's emotions based on their voice tone
    \item AI that drives a vehicle
    \item AI that greets passersby
    \item AI for robotic cleaning or vacuuming
\end{itemize}
This indicates that the IML experience with the device made participants aware of the broader possibilities for AI implementation, leading to various ideas.

Another particularly notable opinion was that development would be feasible if sufficient preparations were made to make it easy to develop. 
For instance, P37 mentioned the AI they could create from now on, ``\textit{Simple ones that are pre-programmed and available as apps (would be possible). I do not think I can create them on my own.}''
P1 also responded, ``\textit{If it is something that can be developed easily like this device,}'' and P4 and P15 similarly mentioned, ``\textit{If it is something simple}''. 
Participants were metacognitively aware that we provided a componentized IML workflow. 
They believed that, at least for experiences like the one they had, they could replicate the small success they achieved.

\subsubsection{Technical Insights and Understanding}
\label{result5}
\new{Based on survey responses, worksheet contents, and interview data, we identified how participants gained technical understanding and developed their own insights. 
We present notable examples from these materials illustrating their learning processes and personal discoveries.}

From the open-ended responses to the questionnaire of ``\textit{What did you learn from the experience?}'', it is clear that participants realized the impact of data on the performance of ML models. 
P26 stated, ``\textit{The quality and quantity of data during the learning process significantly affect performance},'' and P6 mentioned, ``\textit{Increasing the number of recording samples clearly improved accuracy, which made me realize the importance of having a large number of samples.}'' 
More advanced insights included P37's comment, ``\textit{I learned that it is necessary to change the quality of the sound rather than just changing the way the same sound is played}'', and P36's remark, ``\textit{I learned the importance of using reproducible sounds}.''

Observing the types of sounds recorded on the worksheets provides more concrete insights into the participants' ideas.
The ML model in our IML system classifies sounds based on acoustic features, and to facilitate this, we provided participants with various goods to record sounds. 
Observing the participants' recording methods, many different sound sources were assigned to each output action. 
However, there are instances where the recorded sounds did not exhibit significant acoustic differences. 
For example, P1 recorded the words ``\textit{go}'' for the forward action and ``\textit{come back}'' for the backward action, both spoken by the same person. 
These sounds are acoustically similar, which may hinder the model's performance. 
Similarly, P2 assigned ``\textit{shake the maracas strongly}'' and ``\textit{shake the maracas softly}'' to two different actions, which may cause a similar result.

\new{Participants' comments in the interview reveal how they deepened their technical understanding while using the system. 
P1 noted, ``\textit{I was surprised that I could quickly create a model that recognizes voice patterns. I thought machine learning would take much longer, but it worked after just four recordings,}'' indicating their learning about the feasibility of few-shot learning. 
P10 also mentioned the simpleness of the interaction \devicename offers: ``\textit{I thought I would need to write programs, but using voice made it simple.}''
P28 discovered the importance of quality training data through experimentation: ``\textit{When there were gaps in the recording, the accuracy decreased. I learned that providing clean data is crucial for model accuracy.}''
P37 explored different approaches by testing various sound sources: ``\textit{Initially, I tried training with just maracas sounds, but it didn't work well. Then I tried combining different sounds like maracas and tambourine, which was still challenging.}''
These examples demonstrate how participants engaged with \devicename through an experimental approach, forming their own hypotheses, testing them through model training, and drawing conclusions. 
Their insights primarily focused on data quality and the number of recordings needed, reflecting genuine technical understanding.}

\subsubsection{Connecting to Personal Experiences}
\label{result6}

\new{Through interviews with participants, we explored how they connected their personal experiences with their interactions using \devicename. 
These conversations revealed how participants developed their own perspectives by relating the ML experience to their individual contexts.}


Although this experience was a one-time hands-on activity, some participants envisioned integrating IML devices and systems into their daily lives to develop customized AI over time. 
P36, drawing from her parenting experience, mentioned the potential of AI that adapts its approach as children grow: ``\textit{I would like an AI that helps children prepare for school or unpack their bags when they return home by giving specific instructions like 'do A, then do B,' so they can get ready efficiently. 
For example, an AI that sharpens pencils or clearly indicates the next action. 
The idea is similar to that of a puppy for children. 
While dogs are beneficial for education, they are also challenging to take care of. 
An AI that assists children like a puppy would be ideal. 
It would adapt its assistance as the child grows, understanding needs without being told. Such a method of assistance would be wonderful.}''
This statement can be attributed to the device's lifelike qualities and the sense of gradually \textit{nurturing} it by continuously providing data, similar to raising a living creature. 
By adapting the data given to the AI according to the child's growth, the AI's behavior can be adjusted to support the child, like a puppy P36 mentioned continuously.

P31 mentioned the possibility of providing the device with various data over time to meet specific needs: ``\textit{I am excited about the possibility that with continuous training, the device might learn to recognize various differences and nuances, but it is hard to tell from just one day. 
If I could own and feed the device, I could train it to recognize specific sounds and cater to different scenarios.
The potential for the device to understand my detailed needs through training is intriguing. 
Focusing training on what I want it to do in various situations could help the device become specialized for those tasks. 
So, even with the same device, each could behave very differently depending on the training.}''
While we did provide goods to ensure sound diversity, bringing the device into daily life for training could collect more scenario-specific sounds. 
As P31 mentioned, this could enable the creation of AI that addresses one's specific needs more effectively.


In this study, we hypothesized that an IML experience with a cuboid device could reduce users' mental burden. 
This hypothesis seemed particularly relevant for some participants. 
P23 emphasized the importance of reducing the mechanical nature of AI when integrating it into everyday life: ``\textit{It is the same with cleaning robots; I am not good at integrating machines into my life. 
For example, I do not like vacuum cleaners. 
How I can use AI without worrying about its mechanical aspects is very important to me emotionally. 
Although I am interested in AI, it is difficult to incorporate it into my daily life, and it does not recognize my needs if I make mistakes in operation. 
I have an image that I need to read specialized books to make it work, so this experience lowered that threshold for me.}''
This opinion aligns with the earlier discussion about integrating devices into everyday life and providing them with data over time. 
How users perceive AI systems/devices and how comfortably they can integrate them into their lives is crucial for allowing diverse users to experience IML, which relies on collecting varied data.


Some participants linked their daily work to this experience, providing unique insights. 
For example, P1, who had previously worked on ``UI design'', mentioned:
``\textit{I have been involved in projects using AI before. 
Due to technical limitations, users often commented that the AI experience was a bit slow, which made me realize it took time. 
Users might get bored or think it is too troublesome in such cases. 
However, today's quick and clear experience was very enlightening. 
Repeating the process of recording and training the model could encourage users to improve accuracy and continue engaging with the AI.}''

P31 was considering the potential of ``\textit{vibration devices that allow people with hearing impairments to enjoy music},'' linking the experience of creating a sound recognition device to their work. 
She noted: ``\textit{In concert halls, performances are conducted for large audiences, so it is challenging to respond to individual requests. 
There are situations where everyone has to listen quietly, which is a general limitation. 
Therefore, providing a musical experience for people with hearing impairments is not usually possible in typical settings. 
So, I started considering (AI experience) for this purpose.}''
P31 was exploring the possibility of creating AI of sound recognition tailored to the needs of specific users, including the AI personalization mentioned above. 
The experience we provided might have been a minimal yet significant step in this direction.

%% file: IJHCI/07_discussion.tex
\section{Discussion}

\subsection{Key Findings}

\subsubsection{Reducing Mental Burden}

Based on the results (Section \ref{result1} and \ref{result3}), we can conclude that \devicename effectively increases user engagement and provides enjoyment and appeal.
\new{The key elements contributing to this engagement were found to be the device's physical appearance and interaction method, which significantly influenced users' impressions.
The psychological impact of appearance on users is significant, and HCI research has explored various contexts, such as zoomorphic/biomorphic device design~\citep{christiansen2024brings,christiansen2023biomorf} and the role of cuteness in interactions to understand how visual design affects users~\citep{fan2023cuteness,wang2024kawaii}.
While \devicename does not explicitly feature cute faces or animal-like forms, users may perceive life-like qualities in how this small device independently thinks and makes decisions.
The device maintains a simple cubic form without overtly emphasizing biological characteristics, and its decision-making is limited to selecting appropriate responses from a finite set of actions based on sound input.
This minimal design approach appears to be one of the key factors enhancing user engagement by leaving more room for imagination and interpretation.}

The device's ability to express output through physical actions also attracts users. 
The desire to make the device perform the correct actions motivates users to create more accurate models. 
Additionally, creating training data by spatially moving the device offers users an enjoyable experience in terms of the interesting way the device's state changes with movement and the satisfaction of accumulating training data.
This interaction, which requires more effort from users than GUI-based systems, might seem inefficient at first glance. 
However, the fact that users preferred this approach suggests that creating an efficient IML system is not the only factor contributing to engagement. 
The attachment and occasional frustration provided by this ``inconvenience'' paradoxically offer users a more immersive experience.
This insight that the challenges and physical engagement can lead to a more immersive and appealing user experience is a crucial perspective for the future design of IML systems.

On the other hand, it should be noted that the strengths and weaknesses of \devicename compared to conventional GUI-based IML systems can vary significantly depending on the target user (Section \ref{result2}). 
For instance, adults who are proficient in PC operations might find it more convenient to have all operations completed within a GUI, avoiding the need to move objects physically. 
Conversely, a toy-like interface would present a lower technical barrier for children or elderly individuals who are less familiar with PC operations.
Moreover, suppose a user's primary goal is to efficiently improve the accuracy of an ML model with minimal effort. 
In that case, the distinction between GUI and device-based interfaces might be less important, and users might even find the GUI version more efficient. 
For users seeking an entry point to experience ML beyond technical barriers, or those who prefer cute objects, an IML system with a friendly form factor could significantly enhance the quality of their experience.
This diversity in user preferences and needs underscores the importance of considering various interaction modalities when designing IML systems to cater to a wide range of users and use cases.

\new{From our findings, we can identify distinct roles for physical UI and GUI systems when considering the temporal progression of users engaging with ML technology - from initial interest to understanding and problem-solving. 
\devicename received comments like ``\textit{cute and approachable}'' and ``\textit{enjoyable ML model creation experience,}'' suggesting its effectiveness in supporting the very first stage: sparking interest before technical understanding begins.
For subsequent stages of gaining understanding and solving specific problems, GUI systems may be more appropriate. 
As evidenced by GUI's superior ease of use scores, once users develop awareness as ML developers or acquire digital literacy, GUI systems can offer more efficient functionality without the potential cumbersomeness of physical UI.
However, concluding that physical UIs are inferior to GUIs for practical applications would be premature. 
While GUIs allow technically literate users to easily customize ML models and offer better scalability as software implementations, physical UIs may have distinct advantages in certain scenarios. 
For instance, when implementing ML models in hardware, such as IoT devices, having the hardware itself serve as a physical IML system for ML development could be advantageous.
Rather than arguing which approach is more or less useful, we should carefully consider which UI is more suitable based on users' current scenarios and objectives.}


\subsubsection{Stimulating Ideas}

Upon reviewing the specific ideas proposed by the participants in the public event (Section \ref{result4} and \ref{result6}), we found that they extended beyond interactions with the physical device to a wide range of scenarios. 
This diversity of ideas suggests that the spatio-physical interaction of \devicename effectively impressed upon users that creating AI/ML models is not limited to sitting at a desk and programming.
This new perspective likely helped break down the impression of ML as a ``complex, untouchable black box'', potentially leading to the generation of various ideas. 
By providing a novel interaction method that challenges stereotypes, \devicename may have opened up new avenues of thought, particularly beneficial for beginners in the field.
This approach of offering new interactions to dismantle stereotypes appears especially crucial for newcomers to the field, as it can make ML concepts more accessible and less intimidating.
However, since some proposed scenarios were unrelated to the experience itself, it is challenging to definitively attribute all these ideas directly to the experience's content. 
There remains room for additional investigation and discussion on how these diverse ideas were conceived and the extent to which the physical interaction with \devicename influenced this creative process.

A noteworthy observation is the variation in focus among users, despite experiencing the same \devicename.
Even with the same model creation experience, some participants found it important to enjoy the experience itself. 
In contrast, others appeared to aim to deepen their technical understanding, as suggested by their questions about the mechanisms of the ML model and comments on how to improve its accuracy.
Furthermore, examples and opinions related to the creation of training data indicated that some users might focus primarily on trial and error to improve the model's inference accuracy.
To provide a meaningful IML experience from educational and practical perspectives, the same device should ideally have a system design that allows deeper exploration based on individual preferences.
For example, effective designs might include the ability to switch between multiple experience courses by pressing a button or preparing multiple mat patterns that can be used according to different interests.

The diversity of ideas proposed by participants can be attributed to the device's design, which avoids association with specific use-case scenarios.
Typically, methods to elicit ideas from event or study participants include ideation about system personas or brainstorming sessions for idea sharing~\citep{weitz2024explaining,taylor2018everybody}. 
In our study, participants could expand their ideas independently without such explicit techniques.
A key factor was that the device did not limit itself to any specific purpose. 
With IML systems designed for particular scenarios (e.g., art restoration, medical image diagnosis, voice conversion), users would need to think within those specific frameworks.
However, \devicename is primarily for experiencing ML model creation, with its potential applications left open-ended by both the device design and our instructions. 
The appearance and specifications of \devicename remain neutral, avoiding strong associations with specific scenarios or tasks.
This lack of constraints likely allows users to explore a wider range of possibilities during their IML experience.

\subsubsection{Improving Technical Understanding}


The explainability of the model's training and inference processes is crucial for promoting users' technical understanding. 
While our approach of providing a device with minimal physical interaction functions helped users feel more comfortable with the model creation process, it also created a trade-off with technical transparency. 
Some participants expressed interest in understanding \devicename's mechanisms and methods to improve accuracy (Section \ref{result5}). 
However, the device's inherent limitations in presenting information made it more of a black box compared to GUI-based systems.
For example, our classification model relies on acoustic features, but some participants recorded the same sound source for different categories, which would not improve the model's performance. 
If participants had understood that inference was based on acoustic features, they might have assigned different types of sounds to different actions or avoided using sounds with similar acoustic features for multiple actions.
Enhancing the system's explainability could, therefore, improve the quality of training data and users' overall understanding. 
Balancing this need for transparency with the simplicity that makes the device accessible to non-technical users remains a challenge in designing such IML systems.

To improve explainability, introducing visualization techniques like GradCam~\citep{selvaraju2017grad} or LIME~\citep{ribeiro2016should} could be considered.
These techniques enhance explainability in ML and enable participants to engage in more efficient trial-and-error iterations to achieve their goals. 
As a secondary effect, they can deepen participants' understanding of ML technology, potentially leading to more realistic ideas for real-world ML applications.
However, introducing such additional features might complicate the system and reduce user-friendliness. 
Increasing the system's capabilities means increasing the number of elements users need to remember and understand during the IML experience.
This will likely increase their cognitive load and distract them from the more essential aspects of the experience.
When developing IML systems for a diverse range of users, it is crucial to design interactions that balance the system's simplicity with the deepening of technical understanding.

\subsection{Limitation and Future Work}

\subsubsection{Device Implementation}

\devicename primarily serves as a tool for experiencing ML, limiting users to designing models that classify a few categories. 
While we asked participants for ideas on real-world AI/ML applications, the device itself is not designed to solve practical tasks directly. 
This gap may potentially constrain the scope of ideas generated by users.
To address this limitation, we could prepare scenarios aligned with real-world tasks, such as ``training an image recognition model in a robotic vacuum cleaner to avoid fragile objects'' and design IML systems based on these scenarios. 
However, this approach might unintentionally narrow the range of user-generated ideas.
If the goal is to elicit diverse ideas, an IML system with fewer usage restrictions might prove more effective. 
Conversely, if we aim to help users recognize a certain real-world application possibility, introducing a task-specific, problem-solving-oriented IML system could be beneficial.
The choice between these approaches depends on the IML system's specific objectives: fostering broad creativity or focusing on practical applications.

\devicename used in this study simplified the experience of ML model creation, and some participants in the public study were aware that they were engaging in a simplified development process.
As discussed in the results, this experience itself has various significant implications. 
However, users are unable to experience lower-level development. 
For example, it is important to allow non-expert users to experience the actual development process that engineers undergo (e.g., training models through Python coding) from an educational and enlightening perspective. 
Additionally, \devicename does not offer more advanced development steps, such as hyperparameter tuning or modifying model architecture, which could be incorporated into IML systems.
While some users may desire more advanced experiences, others might be satisfied with a basic introduction. 
It is desirable to have a system that can be flexibly adapted to accommodate different user preferences and levels of engagement. 
This approach allows for customization based on individual users' needs and desired experience depth.

\subsubsection{Research Methodology}

\new{Before discussing the specifics of our research methodology, we must acknowledge its inherent limitations.
Firstly, the questionnaire items for quantitative evaluation were custom-developed for this study, which introduces a degree of arbitrariness, as they are not based on standardized usability scales.
Secondly, because participants were allowed to define their own classification tasks, it was impossible to create a standardized test dataset. 
Consequently, a direct, quantitative comparison of the performance of the models created by participants was infeasible and thus excluded from the scope of our evaluation.}

In conducting our study to compare \devicename with its GUI version, we developed a web application that aimed to reproduce the specifications of \devicename faithfully. 
However, determining which aspects should be common and which should differ is arbitrary.
For instance, regarding how \devicename performs actions during inference, one option is to display the action name on the screen. 
Alternatively, we could have implemented a virtual device on the screen that performs the actions to increase fidelity. 
Similarly, for the feature of placing the device on the mat, we can allow users to drag and drop a virtual device onto a simulated mat on the screen or turn each mat square into a clickable button.
There is no clear-cut correct approach here, and it is possible that implementing the GUI version differently could have led to different study results. 
This ambiguity in translating physical interactions to a GUI environment highlights the challenges in creating truly equivalent comparisons between physical and digital interfaces.

For the sake of clarity and ease of understanding for users, we chose the simplest classification model to employ in \devicename. 
This trend is consistent with prior IML research~\citep{talbot2009ensemblematrix,kawabe2024technical,tatsuya2020investigating}, where classification is often selected as the simplest case to implement systems. 
However, the range of ML tasks applicable in the real world is significantly broader. 
For example, regression tasks (e.g., estimating bounding boxes in object recognition from images) and data generation fall under entirely different frameworks than classification. 
How to enable novice users to experience these non-classification tasks remains a challenge. 
Since the method of creating training data also differs, the format of the play mat would not be based on categories for such tasks.
The key to assigning tasks other than classification is to help users understand their specific characteristics. 
For example, object detection involves challenges such as accurately aligning bounding boxes to objects, detecting unintended objects, and detecting differences in the granularity of boxes (e.g., detecting an entire person rather than just their head).
Learning about a task's difficulties and how to overcome them is likely to be valuable when users apply that task to real-world situations.

\new{Last but not least, while we conducted two studies focusing on how \devicename enhances user engagement, there remain unexplored areas of \devicename's potential positive effects on users. 
For instance, we have not thoroughly examined the technical insights and understanding of ML that users acquire through their experience with \devicename. 
Through interviews and ethnographic research, we could investigate what aspects users focus on, what hypotheses guide their device interactions, and what technical understanding they ultimately develop.
Additionally, the findings from the public venue study cannot be interpreted as pure effects of physical UI since they were not directly compared with GUI interactions. 
This could be addressed by re-examining the public venue analyses in the context of GUI comparison, allowing us to isolate the inherent effects of physical UI itself.}

%% file: IJHCI/08_conclusion.tex
\section{Conclusion}

In this paper, we introduced \devicename, a spatio-physical IML device designed to make ML model creation more accessible and engaging for novice users. 
Our comparative study with a GUI version demonstrated that \devicename was more enjoyable and effective at increasing user engagement in the IML process. 
The public venue study further validated these findings, with participants reporting high levels of engagement after interacting with \devicename and generating diverse ideas for ML applications. 
However, we also identified limitations in \devicename's ability to explain certain ML concepts, highlighting the need to further explore IML system UI design. 
These results suggest that while physical, toy-like interfaces can significantly lower barriers to ML engagement for non-experts, there is still room for improvement in balancing hands-on interaction with comprehensive ML understanding. 
Future research should focus on developing IML systems that combine the engaging qualities of physical interfaces with more robust explanatory capabilities. 
These systems aim to provide users with an approachable first step into ML model creation that empowers them to solve their own problems using self-created ML models.